\documentclass[11pt,twoside]{article} 
\usepackage{asp2004}
\usepackage{epsf}
\usepackage{psfig}
\usepackage{lscape} 

\begin{document} 
\title{Discovery of magnetic fields in CPNs}
\author{S. Jordan} 
\affil{ Astronomisches Rechen-Institut, M\"onchhofstr. 12-14, 
    69120 Heidelberg, Germany}
\author{K. Werner} 
\affil{Institut f\"ur Astronomie und Astrophysik, Universit\"at T\"ubingen, Sand 1, 72076 T\"ubingen, Germany
    }
\author{S.J. O'Toole} 
\affil{Dr.-Remeis-Sternwarte Bamberg, Sternwartstr. 7, 96049 Bamberg, Germany
    }
\begin{abstract} 
For the first time we have directly detected  magnetic fields in  central stars of
planetary nebulae by means of  spectro-polarimetry with FORS1 at the VLT. In all four objects of our sample
we found kilogauss magnetic fields, in NGC\,1360 and LSS\,1362  with very high significance, while
in Abell\,36 and EGB\,5 the existence of a  magnetic field is probable but with less certainty.
This discovery supports the hypothesis that the non-spherical symmetry of most
planetary nebulae is caused by  magnetic fields in AGB stars. Our high discovery rate demands mechanisms to prevent full conservation of magnetic flux
during the transition to white dwarfs.
\end{abstract}

\section{Introduction}
The reason why more than 80\%\ of the known planetary nebulae (PNe) are mostly
bipolar and not spherically symmetric
(Zuckerman \&\ Aller 1986, Stanghellini et al. 1993)
is barely understood.
A review on observational and theoretical studies of the shaping of
planetary nebulae is given by Balick \&\ Frank (2002).
It is possible that magnetic fields from the stellar surface are wrapped up by differential rotation so that
the later post-AGB wind will be collimated into two lobes
Garc{\'i}a-Segura et al.\ (1999). Another scenario
says that  magnetic pressure at the stellar surface plays an important role driving the stellar
wind on the AGB (Pascoli 1997).
The idea that  magnetic fields are important has been supported by the detection of polarization in
radio data of circumstellar envelopes of AGB stars
(Kemball \&\ Diamond 1997, Szymczak \&\ Cohen 1997, Vlemmings et al. 2002).
However, until now no magnetic fields have ever directly been detected in central stars of PNe.

\section{Observations and data reduction}
The observations in the spectral range 3400--5900\,\mbox{\AA}
with a spectral resolution
of 4.5\,\mbox{\AA}  of the bright ($V\le 12\fm 5$)
 central stars of NGC\,1360, EGB\,5,
LSS\,1362, and Abell\,36 were obtained
in service mode between
November 2, 2003, and January 27, 2004, with the FORS1 spectrograph
of the UT1 (``Antu'') telescope of the  VLT, which is able to measure
circular polarization with the help of a
 Wollaston prism and rotatable retarder
 plate mosaics in the parallel beam allowing linear and circular polarimetry
 and spectropolarimetry.
NGC\,1360 was observed four times, the other objects only once.
The data reduction was performed analogously to Aznar Cuadrado et al. (2004).
Wavelength calibration is particularly important for this kind of
spectropolarimetric study, and special care was taken to ensure its
accuracy. 

\section{Determination of the  magnetic field strengths}
\begin{figure*}[t]
\begin{minipage}[t]{0.49\textwidth}
\centerline{\psfig{figure=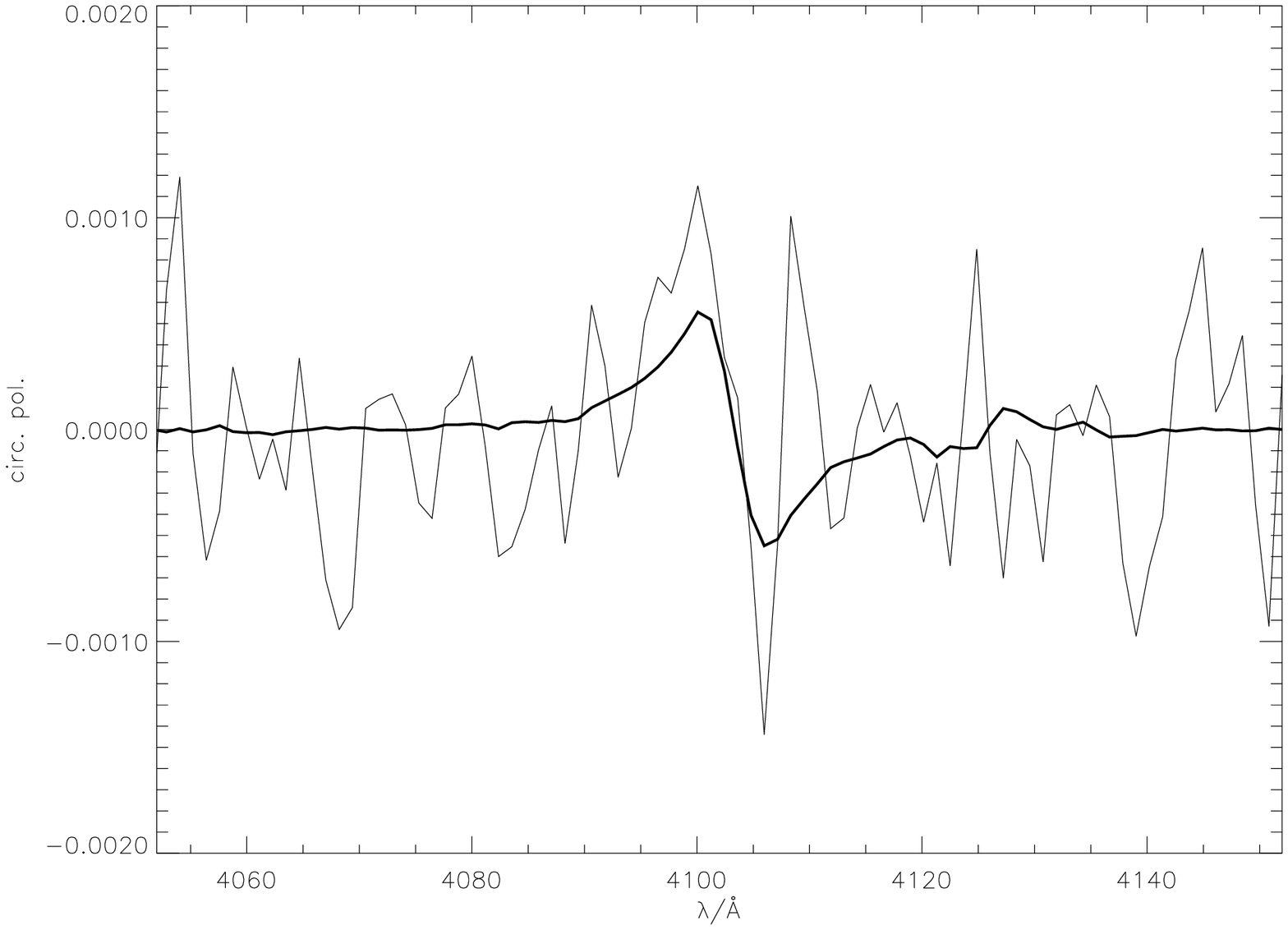,width=\textwidth}}
\end{minipage}
\begin{minipage}[t]{0.49\textwidth}
\centerline{\psfig{figure=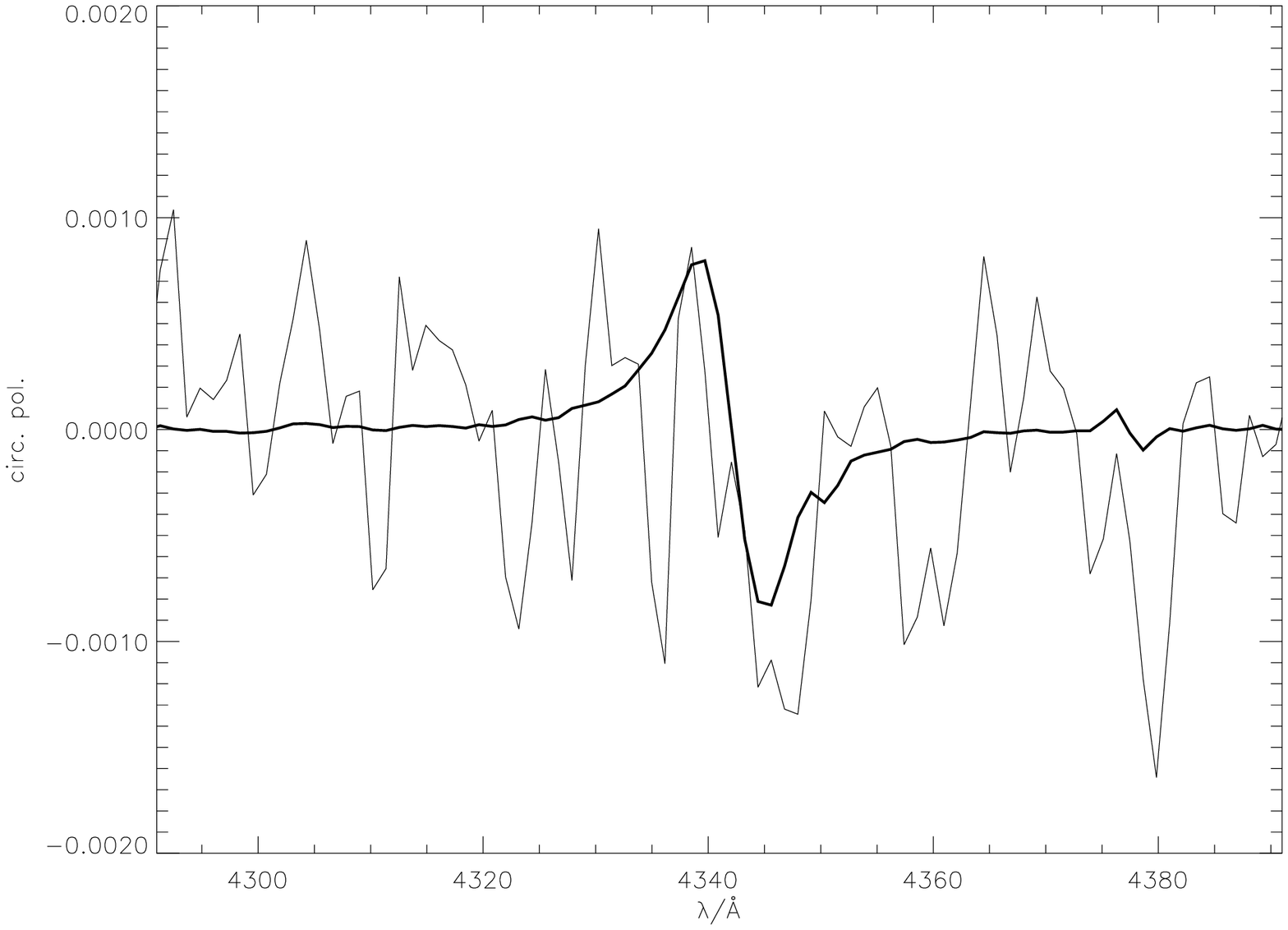,width=\textwidth}}
\end{minipage}
\\
\begin{minipage}[b]{0.49\textwidth}
\centerline{\psfig{figure=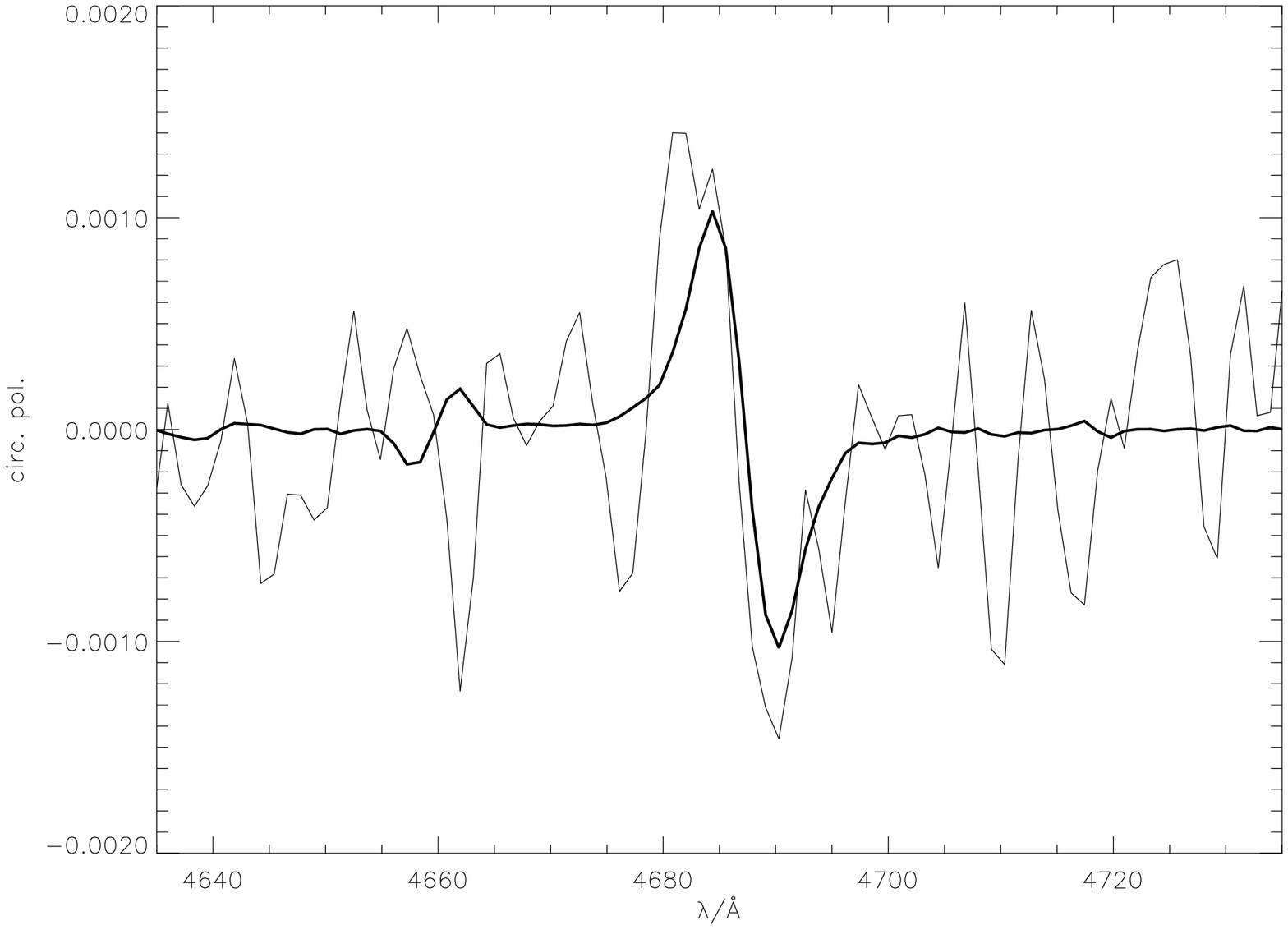,width=\textwidth}}
\end{minipage}
\begin{minipage}[b]{0.49\textwidth}
\centerline{\psfig{figure=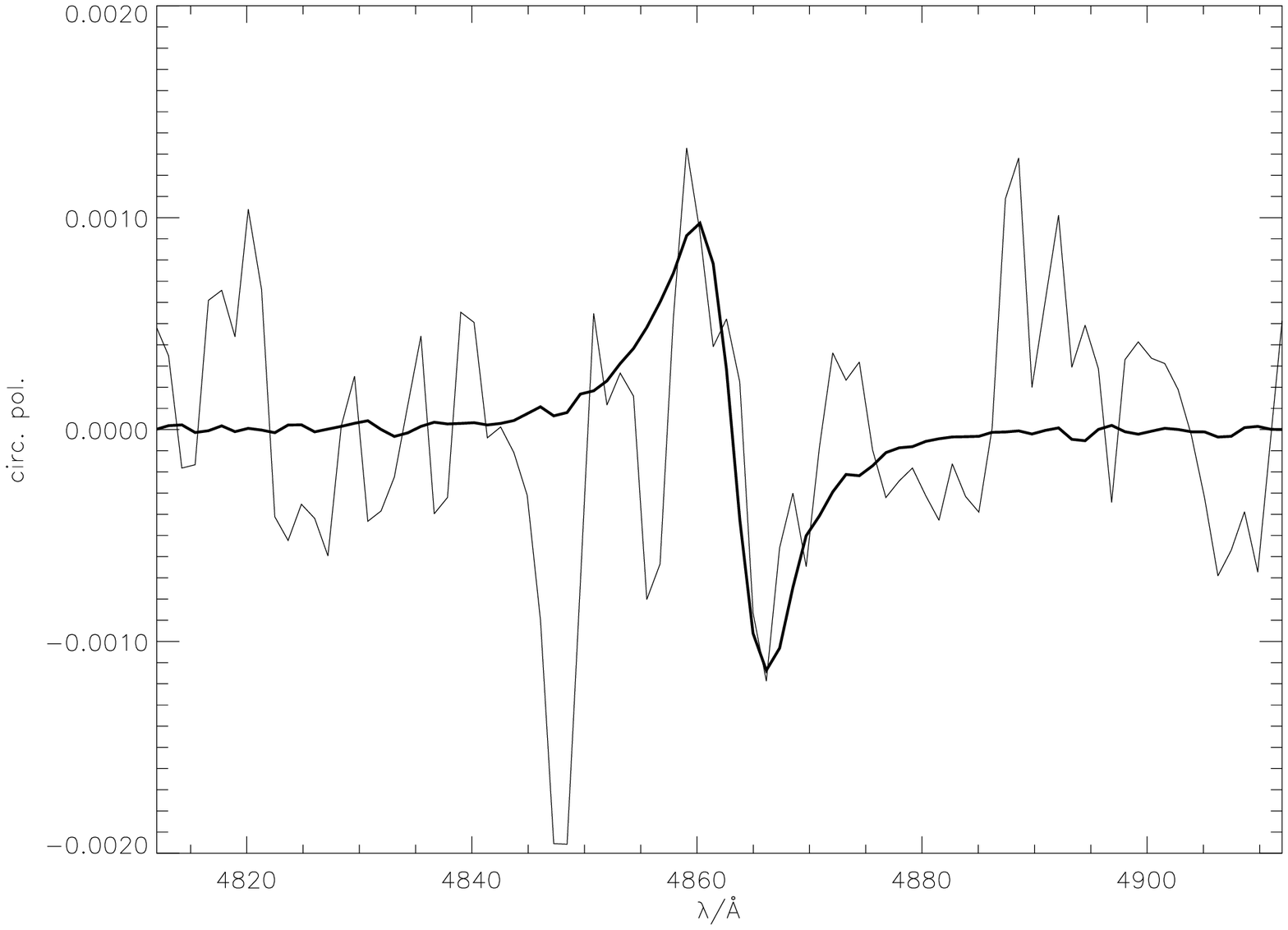,width=\textwidth}}
\end{minipage}
\caption{Circular polarization  ($V/I$) in the third observation
block 
of the central star of NGC\,1360 in the vicinity of the strong
spectral lines H$\delta+$He\,\textsc{ii}, H$\gamma+$He\,\textsc{ii}, He\,\textsc{ii} 4686, H$\beta+$He\,\textsc{ii}
compared to the prediction by the low-field
approximation using a longitudinal magnetic
field of $2832$\,G.}
\end{figure*}

The
longitudinal component of the magnetic field  for
each measurement was determined by comparing
the observed circular polarization for an
interval of $\pm 20$\,\,\mbox{\AA} around the four strongest absorption lines H$\beta+$He\,\textsc{ii},
He\,\textsc{ii} 4686, H$\gamma+$He\,\textsc{ii}, H$\delta+$He\,\textsc{ii}
with the prediction of the weak-field approximation for a given
longitudinal component $B_l$ of the magnetic field as described in
Angel \&\ Landstreet (1970) and Landi \&\ Landi (1973).
The maximum
field strength is in general larger than the longitudinal component.
Test calculations with theoretical spectra have shown that the
blending of the hydrogen lines and the He\,\textsc{ii} lines
introduces uncertainties of the order of 
200\,G.
Only the He\,\textsc{ii} 4686 line is  
not effected by blending.

We determined the statistical errors 
 from the rms
deviation of the observed circular polarization from the best-fit model.
For the central star of NGC\,1360 the weighted mean for the strongest
four spectral lines (H$\delta+$He\,\textsc{ii},H$\gamma+$He\,\textsc{ii},
He\,\textsc{ii} 4686, and H$\beta+$He\,\textsc{ii}) for the four observing
blocks was $B_l/G=-1343\pm 259(\pm 668)$  (68.3\%\ and 99\%\ confidence level),
$1708\pm 257(\pm 664), 2832\pm 269(\pm 695)$, and $194\pm 277(\pm 2548)$.
For EGB\,5 we obtained $1992\pm 562(\pm 1449)$, for LSS\,1362 $1891\pm 371
(\pm 912)$, and for Abell\,36 $1169\pm 466(\pm 1202)$.

NGC\,1360 clearly shows the effect of rotation between
the observations: $-1343, 1708, 2832,$ and $194$\,G.
The  difference in time between the three observations
was 42, 0.8, and 1.0 days. 
The fits to the observations of the third observing block are shown
in Fig.\,1.

We carefully tested our $\chi^2$ fit procedure with several thousand
artificial spectra and found that it is indeed possible
to determine the magnetic field strength even when the magnetic
polarization signal is of the same order of magnitude as the noise.

\section{Discussion and conclusions}
We have detected magnetic fields in 50\%-100\%\ of our small survey
for magnetic fields in central stars of planetary nebulae, depending
on how conservatively  the criteria for statistical significance are
set. This provides very strong support for theories which explain the
non-spherical symmetry (bipolarity) of the majority of planetary
nebulae by magnetic fields. In this first survey we have not performed
a cross check with any spherically-symmetric nebulae, although 
we have submitted a proposal for 
follow-up observations.

Although based on only four objects, our extremely high discovery
rate demands that magnetic flux must be lost during the transition
phase between central stars and white dwarfs: if the magnetic flux
was fully conserved, our four central stars will have fields
between 0.35 and 2\,MG
when they become white dwarfs, deduced from the atmospheric parameters
and the mass-radius relation of Wood (1994).
 Although the number of white dwarfs with
magnetic fields is still a matter of debate, with a range between about 3
and 30\%, even the latter value, which includes
objects with kG field strengths (Aznar Cuadrado et al. 2004), is far
off our high number.
Liebert et al. (2003) quantified the incidence of magnetism at
the level of $\sim$ 2\,MG or greater to be of the order of
 $\sim$10\%.
 This argument would not change by much
if we consider that we have so far only looked at central
stars with non-spherical symmetric nebulae.
An almost 100\%\ probability of magnetic fields larger that
100\,kG can be excluded by the data from the SPY survey 
(Napiwotzki et al. 2003) as
well as the sample from Aznar Cuadrado et al. 2004). It is also worth mentioning that
our central stars have typical white dwarf masses (0.48-0.65\mbox{\,$\rm M_{\odot}$}) and are
not particularly massive.  White dwarfs with MG fields tend to be
more massive than non-magnetic objects (Liebert 1988).

If the magnetic field is located deep in the degenerate
core of the central star, it is very difficult to imagine a
mechanism to destroy the ordered magnetic fields. Therefore,
it would be more plausible to argue that the magnetic field
in the central stars is present mostly in the envelope where it
can be affected by convection and mass-loss. For central stars
hotter than 100\,000\,K we do, however, not expect convection;
only in the central star of EGB\,5 we cannot exclude such a mechanism.

If we assume that the magnetic fields are fossil and magnetic
flux was conserved until the central-star phase, we
estimate that the field strengths on the main sequence
were 9-50\,G, which are not directly detectable. Therefore, our measurement
may indirectly provide evidence for such low magnetic fields on the main
sequence.

Polarimetry with the VLT has led to discovery of magnetic
fields in a large number of objects in the final stage
of stellar evolution: white dwarfs (Aznar Cuadrado et al. 2004),
hot subdwarf stars O'Toole et al. (these proceedings), 
and now in central stars
of planetary nebulae. Although we have now provided a
good basis for the theoretical explanation of the planetary nebula
morphology -- which can more quantitatively be correlated
with additional observations in the future --  new questions about the number
statistics of magnetic fields in the late stages of stellar evolution have
been raised.
The full details of our analysis can be found in Jordan et al. (2005).

\acknowledgements{
We thank the staff of the ESO VLT for carrying out the service
observations. Work on magnetic white dwarfs in T\"ubingen is supported by the
DLR grant 50 OR 0201, SJOT by 
DLR grant 50 OR 0202.}

\end{document}